\documentclass[12pt]{article}

\pdfoutput=1

\usepackage{graphics}
\usepackage{epsfig}
\usepackage{amsfonts}
\usepackage{amssymb}
\usepackage{amsmath}
\usepackage{dsfont}
\usepackage{pifont}
\usepackage{bbm}
\usepackage{multirow}
\usepackage{latexsym}
\usepackage{verbatim}
\usepackage{mcite}
\usepackage{cite}
\usepackage{units}
\usepackage[all]{xy}
\usepackage{cancel}
\usepackage{slashed}

\def\a{{\alpha}}

\newcommand{\be}{\begin{equation}}
\newcommand{\ee}{\end{equation}}
\newcommand{\ben}{\begin{displaymath}}
\newcommand{\een}{\end{displaymath}}
\newcommand{\bea}{\begin{eqnarray}}
\newcommand{\eea}{\end{eqnarray}}

\newcommand{\bean}{\begin{eqnarray*}}
\newcommand{\eean}{\end{eqnarray*}}

\DeclareMathAlphabet{\mathpzc}{OT1}{pzc}{m}{it}

\begin{document}
\pagestyle{plain}


\makeatletter \@addtoreset{equation}{section} \makeatother
\renewcommand{\thesection}{\arabic{section}}
\renewcommand{\theequation}{\thesection.\arabic{equation}}
\renewcommand{\thefootnote}{\arabic{footnote}}


\setcounter{page}{1} \setcounter{footnote}{0}


\begin{titlepage}

\begin{flushright}
UUITP-16/13\\
\end{flushright}

\bigskip

\begin{center}

\vskip 0cm

{\LARGE \bf Accelerated Universes from type IIA\\[2mm] Compactifications} \\[6mm]

\vskip 0.5cm

{\bf Johan Bl{\aa}b\"ack, Ulf Danielsson \,and\, Giuseppe Dibitetto}\\

\vskip 25pt

{\em Institutionen f\"or fysik och astronomi, \\ 
Uppsala Universitet, \\ 
Box 803, SE-751 08 Uppsala, Sweden \\
{\small {\tt \{johan.blaback, ulf.danielsson, giuseppe.dibitetto\}@physics.uu.se}}} \\

\vskip 0.8cm

\end{center}

\vskip 1cm

\begin{center}

{\bf ABSTRACT}\\[3ex]

\begin{minipage}{13cm}
\small

We study slow-roll accelerating cosmologies arising from geometric compactifications of type IIA string theory on $T^{6}/(\mathbb{Z}_{2}\,\times\,\mathbb{Z}_{2})$. 
With the aid of a genetic algorithm, we are able to find quasi-de Sitter backgrounds with both slow-roll parameters of order $0.1$. Furthermore, we study their evolution by numerically solving the 
corresponding time-dependent equations of motion, and we show that they actually display a few e-folds of accelerated expansion.
Finally, we comment on their perturbative reliability.

\end{minipage}

\end{center}

\vfill

\end{titlepage}


\tableofcontents

\section{Introduction}
\label{sec:introduction}

Since the end of last century, various cosmological observations have provided evidence for the existence of dark energy. Combined measurements coming from supernovae 
\cite{Riess:1998cb}, the Cosmic Microwave Background (CMB) radiation \cite{Jaffe:2000tx} and the Baryonic Acoustic Oscillations (BAO) 
\cite{Tegmark:2003ud} have lead to the conclusion that we live in a universe with positive and small cosmological constant. The energy/matter content giving the best fit gave rise to the so-called 
concordance model of cosmology.  

However, despite the increasing precision of the cosmological data released in the past decade by WMAP \cite{Bennett:2012zja}, and most recently by Planck \cite{Ade:2013zuv}, we are still
unable to exclude that such dark energy is described by a quintessence rather than a cosmological constant. Indeed, the current best fit coming from {Planck + WMAP9 + $e$CMB + BAO + SN$e$} for the
$w \, \equiv \, p/\rho$ parameter is \cite{Hazra:2013uqa}
\be\label{eq:expr}
w(a) \, = \, w_{0} \, + \, w_{a} \, (1-a) \ , \, \textrm{ with } \quad \left\{
\begin{array}{cclc}
w_{0} & = & -1.090 \hspace{-1mm}{\tiny\begin{array}{c} + \, 0.168 \\[-1mm] - \, 0.206\end{array}} & , \\[2mm]
w_{a} & = & -0.27 \hspace{-1mm}{\tiny\begin{array}{c} + \, 0.86 \\[-1mm] - \, 0.56\end{array}} & .
\end{array}\right.
\ee

In parallel to this development, it has turned out to be remarkably difficult to construct rigorous models in string theory with perturbatively stable de Sitter (dS) vacua. To find dS solutions one typically needs to introduce ingredients that are not well under control. This includes KKLT \cite{Kachru:2003aw}, where non-perturbative effects play an important role. It is the aim of the present paper to study what happens if you relax the requirement of time independence of the cosmological constant, and instead investigate quintessence scenarios with quasi dS solutions, \emph{i.e.} accelerated expanding solutions which are slowly decaying in time. Apart from serving the purpose of describing the late-time accelerated phase of the universe, this class of 
models has also been considered in the attempt of describing inflation within string theory \cite{Conlon:2005jm, Balasubramanian:2005zx, McAllister:2007bg} and supergravity 
\cite{GomezReino:2006wv, GomezReino:2006dk, GomezReino:2008bi, Covi:2008cn, Achucarro:2012hg, Borghese:2012yu}.

Focusing on flux compactifications of type IIA string theory, one of the first examples of moduli stabilisation at large volume and small string coupling at a classical level can be found in 
ref.~\cite{DeWolfe:2005uu}, where it was achieved by means of a Calabi-Yau compactification with NS-NS 3-form flux, R-R fluxes, D$6$-branes and O$6$-planes. Unfortunately, the corresponding vacua exhibit
negative values of the cosmological constant.

When concentrating on dS solutions in the context of type IIA flux compactifications on a six-torus with D$6$-branes and O$6$-planes, the above setup turns out to fall into a no-go theorem 
\cite{Hertzberg:2007wc} providing a lower bound of $\mathcal{O}(1)$ for the first slow-roll parameter, thus ruling out dS vacua and quasi-dS solutions. 
A possible way to circumvent this result would be to replace the flat six-torus by a negatively curved internal manifold, \emph{i.e.} by adding metric flux.

A lot of work has been done in this context by making use of the underlying minimal supergravity description in four dimensions \cite{Giddings:2001yu,Kachru:2002he, Derendinger:2004jn, DeWolfe:2004ns, 
Camara:2005dc, Villadoro:2005cu, Derendinger:2005ph, Aldazabal:2006up, Aldazabal:2007sn}. However, many indications were found that not only are there no stable dS
vacua in this setup, but also no quasi-dS solutions, at least not in the isotropic sector of the theory. Here, the best one can achieve is a family of unstable dS solutions with second slow-roll 
parameter of $\mathcal{O}(-1)$, which were studied in refs~\cite{Caviezel:2008tf, Danielsson:2009ff, deCarlos:2009fq, Danielsson:2010bc, Dibitetto:2010rg, Danielsson:2011au,
Danielsson:2012et}\footnote{In \cite{Blaback:2013ht} perturbatively stable dS solutions were found provided that non-geometric fluxes were introduced.}.

Very recently solutions that display cosmic acceleration were presented in \cite{Dodelson:2013iba} where non-critical strings were considered, and in \cite{Gur-Ari:2013sba} where discrete Wilson lines among other objects were used.

In the present work, we would like to stress that, by exploring the non-isotropic sectors of the $\mathcal{N}=1$ supergravity theories arising from geometric type IIA compactifications with O$6$/D$6$ on
$T^{6}/(\mathbb{Z}_{2}\,\times\,\mathbb{Z}_{2})$, it is actually possible to find flux backgrounds describing quasi-dS cosmologies with slow-roll parameters both of $\mathcal{O}(.1)$.\footnote{This distinguishes our approach from \cite{Emparan:2003gg} where it is described how acceleration can be achieved beyond a slow-roll regime.} These solutions are hence about $2\sigma$ away from the current cosmological data presented in (\ref{eq:expr}). The method used to find these solutions were by the means of a genetic algorithm which is very similar to the one designed and presented in ref.~\cite{Blaback:2013ht}.

The paper is organised as follows. In section~\ref{sec:review}, we review the details of type IIA flux compactifications $T^{6}/(\mathbb{Z}_{2}\,\times\,\mathbb{Z}_{2})$ with D$6$-branes and O$6$-planes,
and give a description thereof in terms of superpotential deformations of minimal supergravity in four dimensions.
In section~\ref{sec:dynamics}, we introduce the study of the dynamics of a set of scalar fields coupled to gravity in a FRW-like background, and discuss the corresponding equations of motion.
In section~\ref{sec:models}, we first present the explicit flux backgrounds realising slow-roll cosmologies and subsequently, after solving the equations of motions, we analyse their time-dependent 
evolution and show that they exhibit a few e-folds of accelerated expansion. In section~\ref{sec:discussion}, we discuss some related issues such as perturbative control and scale separation.
Finally, we add our conclusions in section~\ref{sec:conclusions}.  

\section{Geometric type IIA compactifications with O$6$/D$6$}
\label{sec:review}

Reductions of type IIA string theory on a twisted $T^{6}$ with fluxes, and one single O$6$-plane, have been extensively studied in the literature. Such orientifold planes split the space-time 
coordinates into transverse and parallel directions as follows 
\be
\begin{array}{lcccc}
\textrm{O}6^{||} \, : &  &  & \underbrace{\times \, \vert \, \times \, \times \, \times}_{D = 4} \,  \, \underbrace{\times \, \times \, \times \, - \, - \, -}_{d = 6} & ,
\end{array}
\notag
\ee
and can be located at the fixed points of the following $\mathbb{Z}_{2}$ involution
\be
\label{O6||_involution}
\begin{array}{lclclc}
\sigma & : & \left(y^{1},\,y^{2},\,y^{3},\,y^{4},\,y^{5},\,y^{6}\right) & \longmapsto & \left(y^{1},\,y^{2},\,y^{3},\,-y^{4},\,-y^{5},\,-y^{6}\right) & ,
\end{array}
\ee
where $\left\{y^{i}\right\}$ denote the six-dimensional compact coordinates.

According to \eqref{O6||_involution}, all the fields and fluxes arising from such a compactification undergo a discrete truncation only retaining parity-allowed objects. This gives rise to an 
effective four-dimensional supergravity theory with $16$ supercharges. In what follows we will see how to use the orbifold symmetry in order to reduce the amount of supersymmetry of the resulting
supergravity down to $\mathcal{N}=1$ in four dimensions.

\subsection*{The $\mathbb{Z}_{2}\,\times\,\mathbb{Z}_{2}$ orbifold}
\label{subsec:Z2Z2}

Let us now consider the $\mathbb{Z}_{2}\,\times\,\mathbb{Z}_{2}$ group given by $\left\{1, \, \theta_{1}, \, \theta_{2}, \, \theta_{1}\theta_{2}\right\}$ generated by 
\be
\begin{array}{lclclc}
\theta_{1} & : & \left(y^{1},\,y^{2},\,y^{3},\,y^{4},\,y^{5},\,y^{6}\right) & \longmapsto & \left(-y^{1},\,-y^{2},\,y^{3},\,-y^{4},\,-y^{5},\,y^{6}\right) & , \\[1mm]
\theta_{2} & : & \left(y^{1},\,y^{2},\,y^{3},\,y^{4},\,y^{5},\,y^{6}\right) & \longmapsto & \left(-y^{1},\,y^{2},\,-y^{3},\,-y^{4},\,y^{5},\,-y^{6}\right) & . 
\end{array}
\ee
At the fixed points of $\left\{\sigma\theta_{1}, \, \sigma\theta_{2}, \, \sigma\theta_{1}\theta_{2}\right\}$ one can locate a triplet of O$6$-planes placed as 
\be
\begin{array}{lccccc}
\textrm{O}6^{\perp}  \, : &  &  & \underbrace{\times \, \vert \, \times \, \times \, \times}_{D = 4} \, &  \, \underbrace{\left\{\begin{array}{c}
- \, - \, \times \, \times \, \times \, - \\[1mm]
- \, \times \, - \, \times \, - \, \times \\[1mm]
\times \, - \, - \, - \, \times \, \times  
\end{array}\right.}_{d = 6} & .
\end{array}
\notag
\ee
Dividing out by this extra discrete orbifold symmetry further truncates the theory to a minimal supergravity in four dimensions.

Orbifold compactifications of type IIA string theory on $T^{6}/(\mathbb{Z}_{2}\,\times\,\mathbb{Z}_{2})$, with O$6$-planes (and duals thereof) and generalised fluxes, can all be placed within the same framework
as the four-dimensional supergravity models that are known as $STU$-models. These theories enjoy $\mathcal{N}=1$ supersymmetry and $\textrm{SL}(2)^{7}$ global bosonic symmetry.
The action of such a global symmetry on the fields and couplings can be interpreted as the effect of string dualities. 

The scalar sector contains seven complex fields spanning the coset space $\left(\textrm{SL}(2)/\textrm{SO}(2)\right)^{7}$, which we denote by $\Phi^{\alpha}\,\equiv\,\left(S,T_{i},U_{i}\right)$ with $i=1,2,3$. 
The kinetic Lagrangian follows from the K\"ahler potential
\be
\label{Kaehler_STU}
K\,=\,-\log\left(-i\,(S-\overline{S})\right)\,-\,\sum_{i=1}^{3}{\log\left(-i\,(T_{i}-\overline{T}_{i})\right)}\,-\,\sum_{i=1}^{3}{\log\left(-i\,(U_{i}-\overline{U}_{i})\right)}\ .
\ee
This yields
\be
\mathcal{L}_{\textrm{kin}} = \frac{\partial
S\partial \overline{S}}{\left(-i(S-\overline{S})\right)^2} + \, \sum_{i=1}^{3}\left(\frac{\partial
T_{i}\partial \overline{T}_{i}}{\left(-i(T_{i}-\overline{T}_{i})\right)^2} + \frac{\partial
U_{i}\partial \overline{U}_{i}}{\left(-i(U_{i}-\overline{U}_{i})\right)^2}\right) \ .
\ee

The presence of fluxes induces a scalar potential $V$ for the moduli fields, which is given in terms of the above K\"ahler potential and a holomorphic superpotential $W$ by
\be
\label{V_N=1}
V\,=\,e^{K}\left(-3\,|W|^{2}\,+\,K^{\alpha\bar{\beta}}\,D_{\alpha}W\,D_{\bar{\beta}}\overline{W}\right)\ ,
\ee
where $K^{\alpha\bar{\beta}}$ is the inverse K\"ahler metric and $D_{\alpha}$ denotes the K\"ahler-covariant derivative.

The general form of a superpotential induced by geometric fluxes in type IIA with O$6$-planes is given by
\be
\label{W_Geom}
W\,=\underbrace{\,P_{1}(U_{i})\,}_{\textrm{R-R fluxes}}\,+\,\underbrace{\,S\,P_{2}(U_{i}) \,+\,\sum\limits_{k}{T_{k}\,P_{3}^{(k)}(U_{i})\,}}_{\textrm{NS-NS fluxes}} \ ,
\ee
where $P_{1}$, $P_{2}$ and $P_{3}^{(k)}$ are (up to) cubic polynomials in the complex structure moduli given by
\be
\begin{array}{cclc}
P_{1}(U_{i}) & = & a_{0}\,-\,\sum\limits_{i}{a_{1}^{(i)}\,U_{i}}\,+\,\sum\limits_{i}{a_{2}^{(i)}\,\dfrac{U_{1}\,U_{2}\,U_{3}}{U_{i}}}\,-\,a_{3}\,U_{1}\,U_{2}\,U_{3} & , \\[3mm]
P_{2}(U_{i}) & = & -b_{0}\,+\,\sum\limits_{i}{b_{1}^{(i)}\,U_{i}} & , \\[3mm]
P_{3}^{(k)}(U_{i}) & = & c_{0}^{(k)}\,+\,\sum\limits_{i}{c_{1}^{(ik)}\,U_{i}} & .
\end{array}
\ee
The IIA flux interpretation of the above superpotential couplings is summarised in table~\ref{table:fluxes}, from which one can see that the total parameter space of geometric fluxes in this duality 
frame consists of $24$ superpotential couplings. However, in order to have a twisted torus interpretation, one needs to impose the following Jacobi constraints
\be
\label{Jacobi}
{\omega_{[AB}}^{E} \, {\omega_{C]E}}^{D} \, = \, 0 \ ,
\ee
where $A,\, B,\dots$ are fundamental $\textrm{SL}(6)$ indices, which are split into $\{a, \, m\}$ by the orientifold involution. This restricts the number of independent metric flux parameters from
$3 \, + \, 9 \, = \, 12$ down to\footnote{This is actually true only in the semisimple branch of solutions of the \eqref{Jacobi}. There are other non-semisimple branches enjoying a smaller parameter space, 
but they look less promising for the aim of finding accelerated solutions, at least according to our indications.} $6$. 

\begin{table}[h!]
\renewcommand{\arraystretch}{1.25}
\begin{center}
\scalebox{0.92}[0.92]{
\begin{tabular}{ | c || c | c | c |}
\hline
couplings & Type IIA & fluxes & \emph{dof}'s\\
\hline
\hline
$1 $ &  $F_{ambncp}$ & $  a_0 $ & $1$\\
\hline
$U_{i} $ &  $F_{ambn}$ & $   -a_1^{(i)} $ & $3$\\
\hline
$U_{j}U_{k} $ &  $F_{am}$ & $  a_2^{(i)} $ & $3$\\
\hline
$U_{i}U_{j}U_{k} $ & $F_{0}$ & $  -a_3 $ & $1$\\
\hline
\hline
$S $& $ {H}_{mnp} $ & $  -b_0$ & $1$\\
\hline
$S \, U_{i} $ &  ${{\omega}_{mn}}^{c}$ & $  b_1^{(i)} $ & $3$\\
\hline
\hline
$T_{i} $& $ H_{a b p} $ & $  c_0^{(i)} $ & $3$\\
\hline
$T_{i} \, U_{j} $ &  $ {\omega_{p a}}^{n} = {\omega_{b p}}^{m} \,\,\,,\,\,\, {\omega_{b c}}^a $  & $c_1^{(ji)} $ & $9$\\
\hline
\end{tabular}
}
\end{center}
\caption{{\it Mapping between fluxes and couplings in the superpotential in type IIA with O$6$-planes. The six internal directions of $T^{6}$ are split into $\,``-"$ labelled by $a=1,2,3$, and 
$\,``\,|\,"$ labelled by $m=4,5,6$ (\emph{i.e.} parallel and transverse to $\textrm{O}6^{||}$ respectively). 
Note that the orbifold involution forces $i,j,k$ to be all different any time they appear as indices of fields of the same type ($T$ or $U$).}}
\label{table:fluxes}
\end{table}

\section{Analysis of the scalar dynamics}
\label{sec:dynamics}

In this section, we will analyse the dynamics of the fourteen scalars generically obtained from the class of type IIA backgrounds described in section~\ref{sec:review} coupled to gravity. This will lead 
us to the derivation of a system of coupled differential equations, which generalises what is normally obtained in the case of single-field inflation.

The sector of the four-dimensional supergravity theory representing the coupling of scalars with gravity is described by the following Lagrangian  
\be
\label{L1}
\mathcal{L} \, = \, \sqrt{-g} \, \left(-\frac{1}{2} \, K_{IJ}(\phi) \, \partial_{\mu}\phi^{I} \, \partial^{\mu}\phi^{J} \, - \, V(\phi)\right) \ ,
\ee
where $I=1, \cdots, \, 14$ and $K_{IJ}$ is derived from the K\"ahler metric $K_{\alpha\bar{\beta}}$ when rewriting the seven complex fields introduced in section~\ref{sec:review} in terms of their real 
degrees of freedom according to
\be
\left\{\begin{array}{lclc}
S & = & \chi \, + \, i \, e^{-\phi} & , \\[1mm]
T_{i} & = & \chi^{(1)}_{i} \, + \, i \, e^{-\phi^{(1)}_{i}} & , \\[1mm]
U_{i} & = & \chi^{(2)}_{i} \, + \, i \, e^{-\phi^{(2)}_{i}} & , 
\end{array}\right.
\ee
where $\left\{\phi^{I}\right\} \, \equiv \, \left\{\phi, \, \phi^{(1)}_{i}, \, \phi^{(2)}_{i}, \, \chi, \, \chi^{(1)}_{i}, \, \chi^{(2)}_{i}\right\}$, with $i=1,2,3$.

On an FRW-like background with the metric defined as 
\be
\textrm{d}s^{2}_{\textrm{FRW}} \, = \, -\textrm{d}t^{2} \, + \, a(t)^{2} \, \left(\textrm{d}x^{2} \, + \, \textrm{d}y^{2} \, + \, \textrm{d}z^{2}\right) ,
\ee
and after choosing a time-dependent profile for all the scalars, one can rewrite the Lagrangian \eqref{L1} as
\be
\label{L2}
\mathcal{L} \, = \, a(t)^{3} \, \left(\frac{1}{2} \, K_{IJ}(\phi(t)) \, \dot{\phi}^{I}(t) \, \dot{\phi}^{J}(t) \, - \, V(\phi(t))\right) \ .
\ee
The corresponding Euler-Lagrange equations read
\be
\label{EOM-phi}
\ddot{\phi}_{I} \, + \, 3\,H\,\dot{\phi}_{I} \, + \, \partial_{K}K_{IJ}\,\dot{\phi}^{K}\,\dot{\phi}^{J} \, + \, \partial_{I}V \, = \, 0 \ ,
\ee
where we have introduced the Hubble parameter $H \, \equiv \, \frac{\dot{a}}{a}$.

In order to write down the Einstein equations describing the dynamics of the metric we need to introduce the energy density $\rho$ and the pressure $p$ parametrising the stress-energy tensor in the rest
frame
\be
\label{rho-p}
\begin{array}{lclc}
\rho & = & \frac{1}{2} \, K_{IJ} \, \dot{\phi}^{I} \, \dot{\phi}^{J} \, + \, V(\phi) & , \\[2mm]
p & = & \frac{1}{2} \, K_{IJ} \, \dot{\phi}^{I} \, \dot{\phi}^{J} \, - \, V(\phi) & .
\end{array}
\ee
In terms of the above quantities, one can write the Einstein equations in the form of the Friedmann equations\footnote{Note that in order to be consistent with the supergravity potential
given in \eqref{V_N=1} we will have to set the reduced Planck mass equal to $1$, \emph{i.e.} $M_{\textrm{Pl}}^{2} \, = \, \frac{1}{8 \pi G} \, = \, 1$.}
\be
\begin{array}{rclc}
\rho \, - \, \dfrac{3 H^{2}}{8 \pi G} & = & 0  & , \\[2mm]
\dfrac{\ddot{a}}{a} \, + \, \dfrac{4}{3} \, \pi \, G \, (\rho \, + \, 3p) & = & 0 & ,
\end{array}
\ee
where the first equation can be essentially interpreted as a Hamiltonian constraint while the second actually describes the dynamics of the scale factor $a(t)$. 

Moreover, the stress-energy tensor defined in \eqref{rho-p} should obey a conservation law which can be written in the form of the continuity equation
\be
\label{continuity}
\dot{\rho} \, + \, 3H \, (\rho \, + \, p) \, = \, 0 \ .
\ee
Such a condition happens to be already implied by the equations of motion for the scalars \eqref{EOM-phi} together with the first Friedmann equation. Moreover, by using this Hamiltonian constraint one more time 
inside \eqref{continuity}, one can derive the second Friedmann equation.

Finally this means that the full set of equations describing the dynamics of the $14$ scalars coupled to FRW gravity is given by
\be
\label{EOMs}
\left\{
\begin{array}{rclc}
\dfrac{1}{2} \, K_{IJ} \, \dot{\phi}^{I} \, \dot{\phi}^{J} \, + \, V(\phi) \, - \, \dfrac{3 H^{2}}{8 \pi G} & = & 0  & , \\[2mm]
\ddot{\phi}_{I} \, + \, 3\,H\,\dot{\phi}_{I} \, + \, \partial_{K}K_{IJ}\,\dot{\phi}^{K}\,\dot{\phi}^{J} \, + \, \partial_{I}V & = & 0  & ,
\end{array}
\right.
\ee
which consists of $15$ coupled differential equations in the $15$ unknown functions $\left\{a(t), \, \phi^{I}(t)\right\}$, appearing up to first and second order respectively.

\subsection*{Slow-roll accelerated expansion}

As we have briefly discussed in the introduction, both inflation and dark energy require a (quasi-)dS phase describing a universe in accelerated expansion. The preferred regime in which one can 
construct such a cosmological model is the so-called \emph{slow-roll} approximation. This essentially corresponds to a situation where the kinetic energy inside \eqref{rho-p} is negligible w.r.t. the 
potential term $V(\phi)$. This already intuitively results in scalar fields which vary quite slowly in time while the scale factor expands exponentially.

Explicitely, the validity of the slow-roll approximation is encoded in the first and second slow-roll parameters
\be
\begin{array}{lclclclc}
\epsilon & \equiv & \frac{3}{2} \, \frac{\dot{\phi}^{2}}{V \, + \, \dot{\phi}^{2}} & \textrm{ and } & \eta & \equiv & -\frac{\ddot{\phi}}{H \, \dot{\phi}} & ,
\end{array}
\ee
in the following way: $\epsilon \, \ll \, 1$ and $|\eta| \, \ll \, 1$. Generically these slow-roll conditions can be translated into properties of flatness of the scalar potential through
\be
\label{slow-roll_V}
\begin{array}{lclclclc}
\epsilon_{V} & \equiv & \frac{1}{2} \, K^{IJ} \, \frac{D_{I}V \, D_{J}V}{V^{2}} \, \ll \, 1 & \textrm{ and } & |\eta_{V}| & \equiv &  
\left|\textrm{Min.~Eig.}\left(\frac{K^{JK} \, D_{I} \, D_{k}V}{V}\right)\right| \, \ll \, 1 & .
\end{array}
\ee
However, whenever such an accelerated expanding phase is driven by more scalars, the conditions \eqref{slow-roll_V} are not sufficient for guaranteeing the existence of such a phase. The typical problem
that one can run into in this case is an $\epsilon_{V}$ parameter mainly generated by a scalar with a big and positive mass combined with the presence of a nearly-flat tachyon driving $\eta_{V}$.
In order to overcome this generic problem, one needs to require that \emph{all} the directions of the scalar manifold significantly contributing to $\epsilon_{V}$ have a nearly-flat normalised mass.
In particular, if one defines the direction of rolling as
\be
\label{rolling_dir}
\phi_{\epsilon}^{I} \, \equiv \, K^{IJ} \, \frac{D_{J}V}{|DV|} \ ,
\ee
and the projection of the mass matrix along such a direction as
\be
\eta_{\textrm{proj.}} \, \equiv \, \frac{\phi_{\epsilon}^{I} \, D_{I}D_{J}V \, \phi_{\epsilon}^{J}}{V} \ ,
\ee
one would also desire $|\eta_{\textrm{proj.}}|$ to be as small as possible.


In the next section we will present some particularly interesting type IIA string theory backgrounds which have values for $\epsilon_{V}$, $\eta_{V}$ and $\eta_{\textrm{proj.}}$ all about $\mathcal{O}(.1)$ or slightly smaller. After discussing some issues related to the choice of suitable initial conditions, we will then solve the system of differential equations given 
in \eqref{EOMs} with the aid of numerical methods and show that they actually provide some e-folds of accelerated expansion.

\section{The explicit accelerated models}
\label{sec:models}

Our genetic algorithm, trying to minimise $\epsilon_{V}$, $|\eta_{V}|$ and $|\eta_{\textrm{proj.}}|$ while keeping $V \, > \, 0$, by varying the values of the fluxes, produced the flux backgrounds presented in 
tables~\ref{table:Sol_1&2}--\ref{table:Sol_3&4} in Appendix \ref{sec:app}. These are not the only backgrounds found but only the four best ones found. Their physical parameters being reported in table~\ref{table:eps_eta}. 


\begin{table}[h!]
\renewcommand{\arraystretch}{1.25}
\begin{center}
\scalebox{0.92}[0.92]{
\begin{tabular}{ | c || c | c | c | c | c |}
\hline
 ID & $\tilde{\gamma}$ & $\epsilon_{V}$ & $\eta_{V}$ & $\eta_{\textrm{proj.}}$ & $\eta_{\textrm{sG}}$ \\
\hline \hline 
$1$ & $1.13633$ & $0.430423$ & $-0.151163$ & $0.0405875$ & 2.1178\\ 
$2$ & $1.19125$ & $0.452063$ & $-0.0776699$ & $0.35292$ & 1.59615\\ 
$3$ & $1.11877$ & $0.383055$ & $-0.162635$ & $0.0251493$ & 2.64077\\ 
$4$ & $1.24603$ & $0.391704$ & $-0.318953$ & $0.00224945$ & 1.44491\\ 
\hline
\end{tabular}
}
\end{center}
\caption{{\it Values of the normalised energy, first and second slow-roll parameters, and the projection of the mass matrix along the direction of rolling for the four best solutions which were found. 
Here we adopt the definition $\tilde{\gamma}\,\equiv\,\frac{|DW|^{2}}{3|W|^{2}}$ given in ref.~\protect\cite{Danielsson:2012by}. The values in the last column represent the averaged sGoldstino mass
as defined in ref.~\protect\cite{Covi:2008cn}. In none of the cases the supersymmetry breaking scalars seem to significantly contribute to the dynamical process of accelerated expansion.}}
\label{table:eps_eta}
\end{table}

\subsection*{Choosing the initial conditions}

Since the equations of motion introduced in \eqref{EOMs} are first order in the scale factor and second order in each of the scalar fields, we need to assign the values of $\left\{a(0), \, \phi^{I}(0),
 \, \dot{\phi}^{I}(0)\right\}$ in order to solve them. By convention, one can always choose $a(0) \, = \, 1$, whereas the scalar fields in our solutions are already sitting at the origin of moduli space, \emph{i.e.} $\phi^{I}(0) \, = \, 0$ for
any $I$. Thus, the only thing that needs to be fixed carefully is the first time-derivative of the scalars at the initial time.

Since we are looking for a slow-roll accelerated dynamics, we have chosen these velocities to be exactly those of slow-roll. This is achieved by
imposing the conditions in \eqref{EOMs} at $t=0$ in the absence of the terms with $\ddot{\phi}$ and $\dot{\phi}^{2}$, \emph{i.e.} 
\be
\left\{
\begin{array}{rccl}
V_{0} \, - \, 3\,H_{0}^{2} & = & 0 & , \\
3\,H_{0} \, \dot{\phi}_{I}(0) \, + \, \partial_{I}V_{0} & = & 0 & ,
\end{array}\right.
\ee
This yields 
\be
\dot{\phi}^{I}(0) \, = \, -\left(\sqrt{3V} \, K^{IJ} \, \partial_{J}V \right)|_{\phi=0} \ .
\ee
Clearly, this choice of initial condition is only possible when $V$ is positive in the origin.

\subsection*{Time evolution and plots}

\begin{figure}
\begin{center}
\scalebox{1}[0.95]{
\begin{tabular}{cccc}
\includegraphics[scale=0.45,keepaspectratio=true]{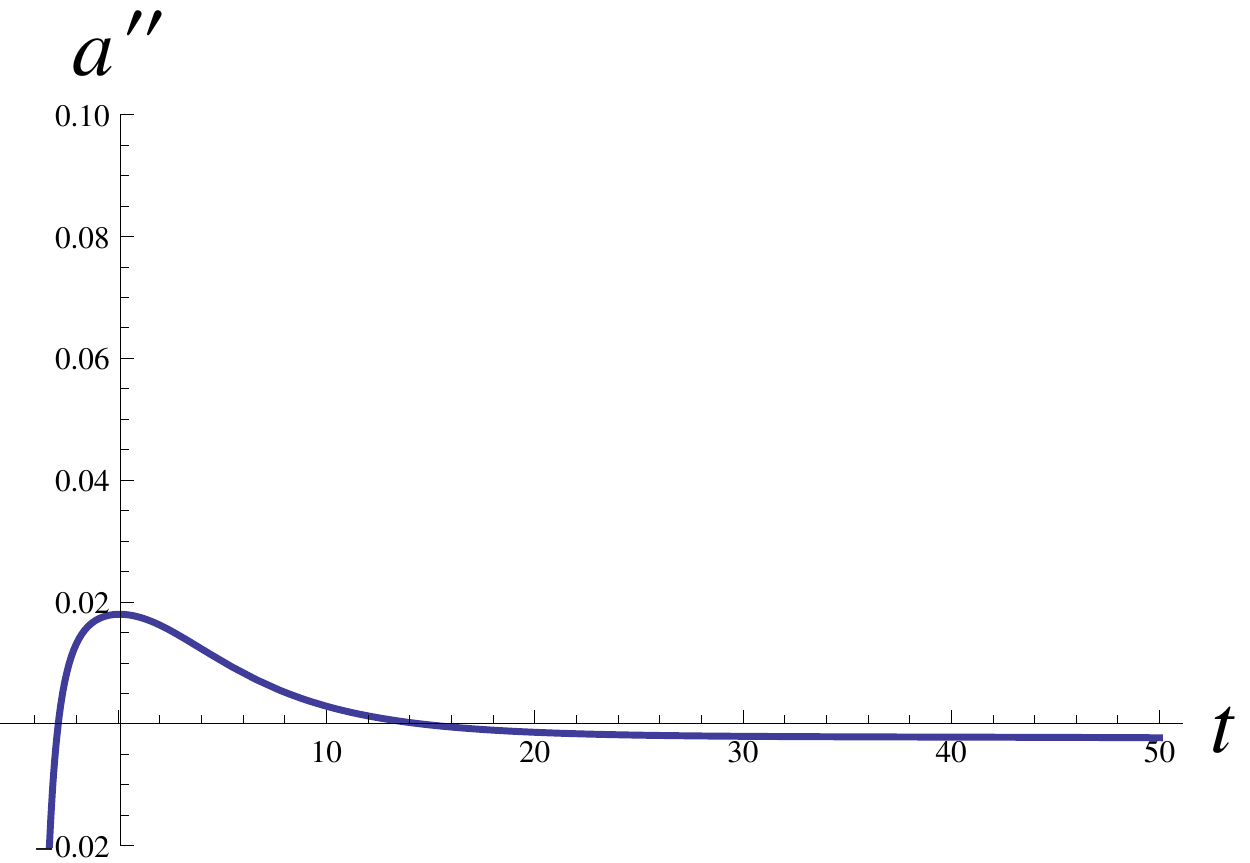} &  &  & \includegraphics[scale=0.45,keepaspectratio=true]{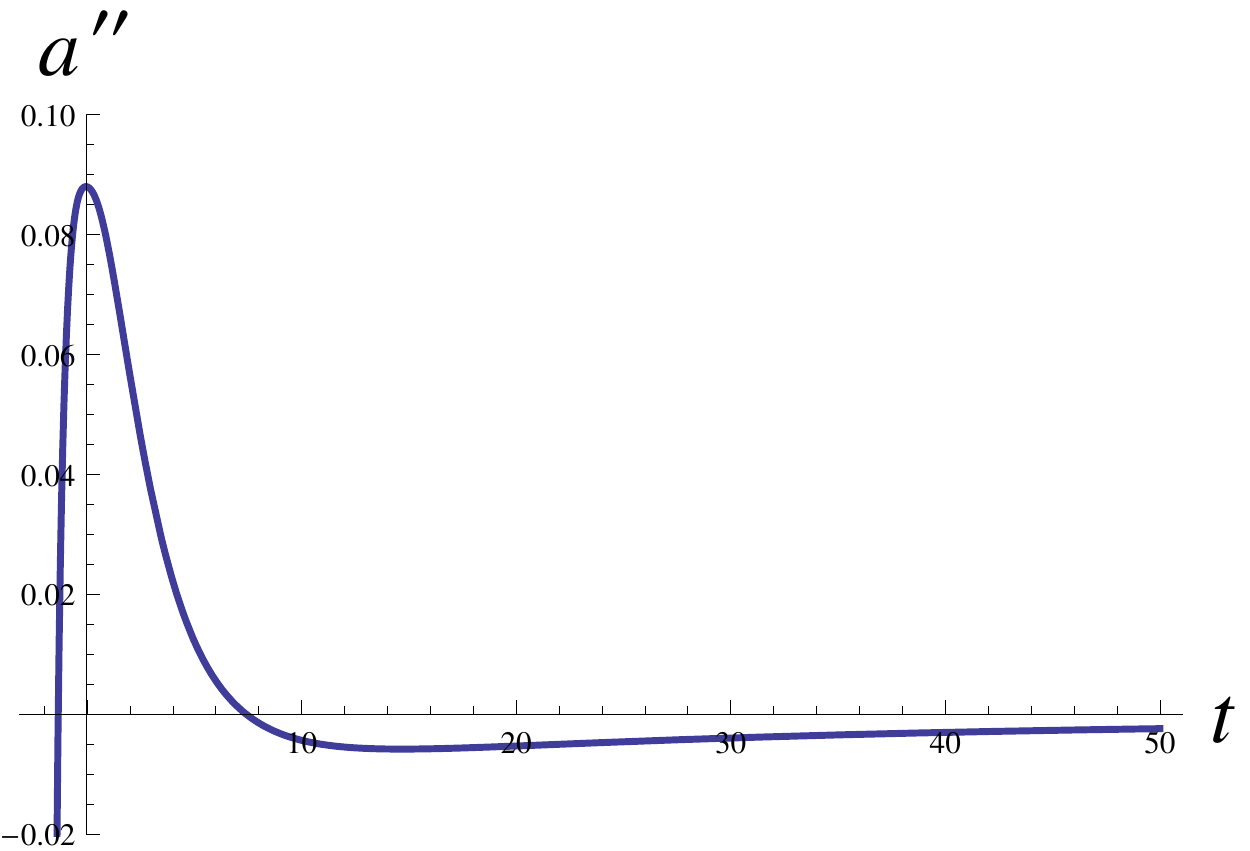}  \\
\includegraphics[scale=0.45,keepaspectratio=true]{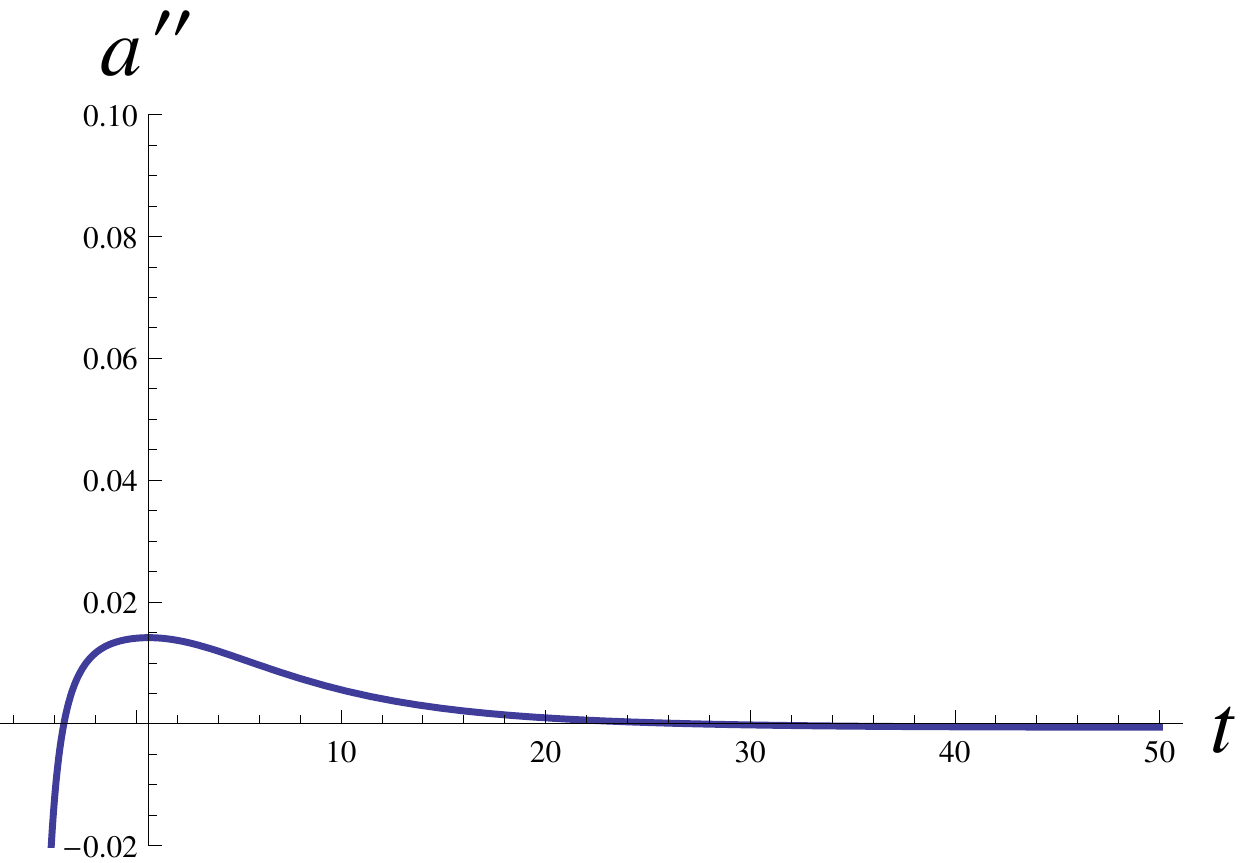} &  &  & \includegraphics[scale=0.45,keepaspectratio=true]{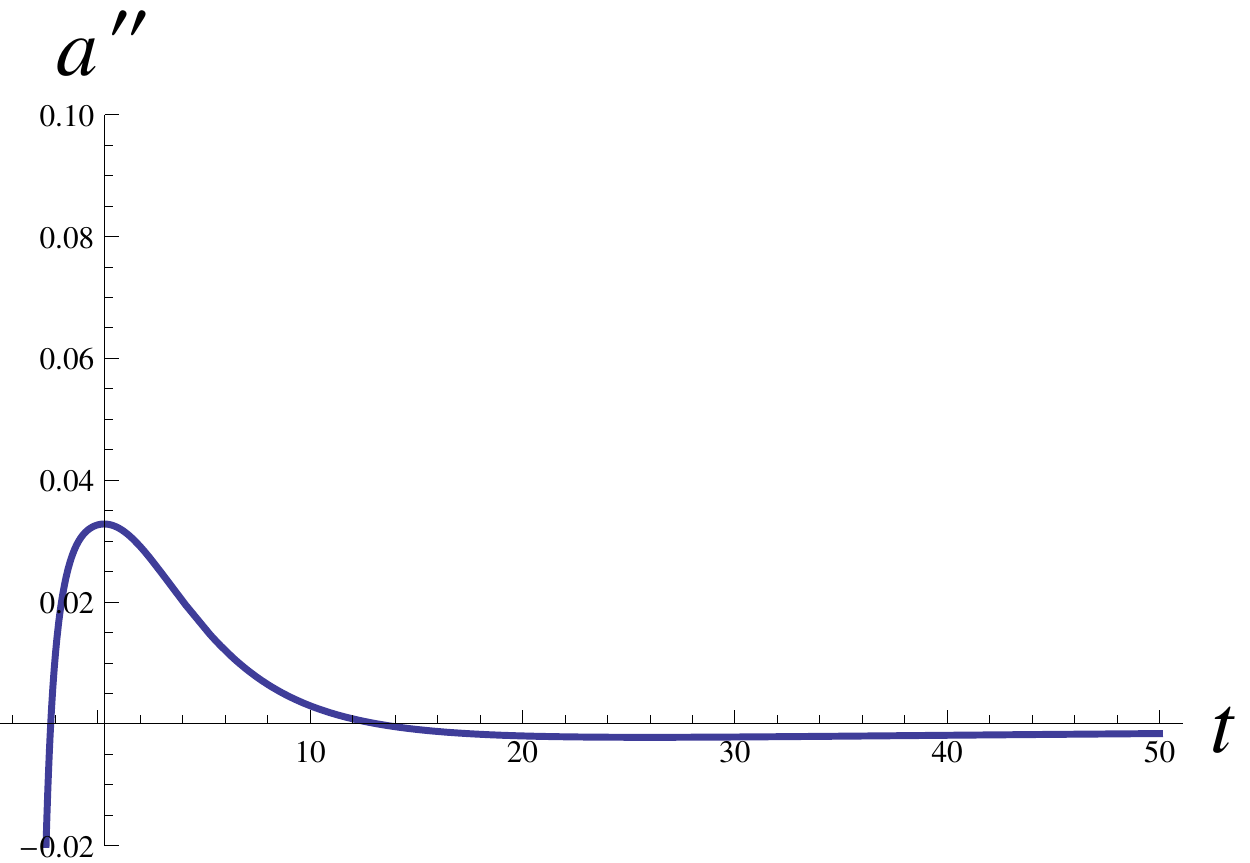}  
\end{tabular}
}
\end{center}
\caption{{\it The second time-derivative of the scale factor $\ddot{a}(t)$ as a function of time for all the different solutions ordered by rows (Sol.~$1$ and $2$ in the first row, Sol.~$3$ and $4$
in the second one). As one can see in all cases, such acceleration turns out to be positive around the time of maximal acceleration (not located at $t=0$).}}
\label{Plots_addot}
\end{figure}
\begin{figure}
\begin{center}
\scalebox{0.97}[0.93]{
\begin{tabular}{cccc}
\includegraphics[scale=0.5,keepaspectratio=true]{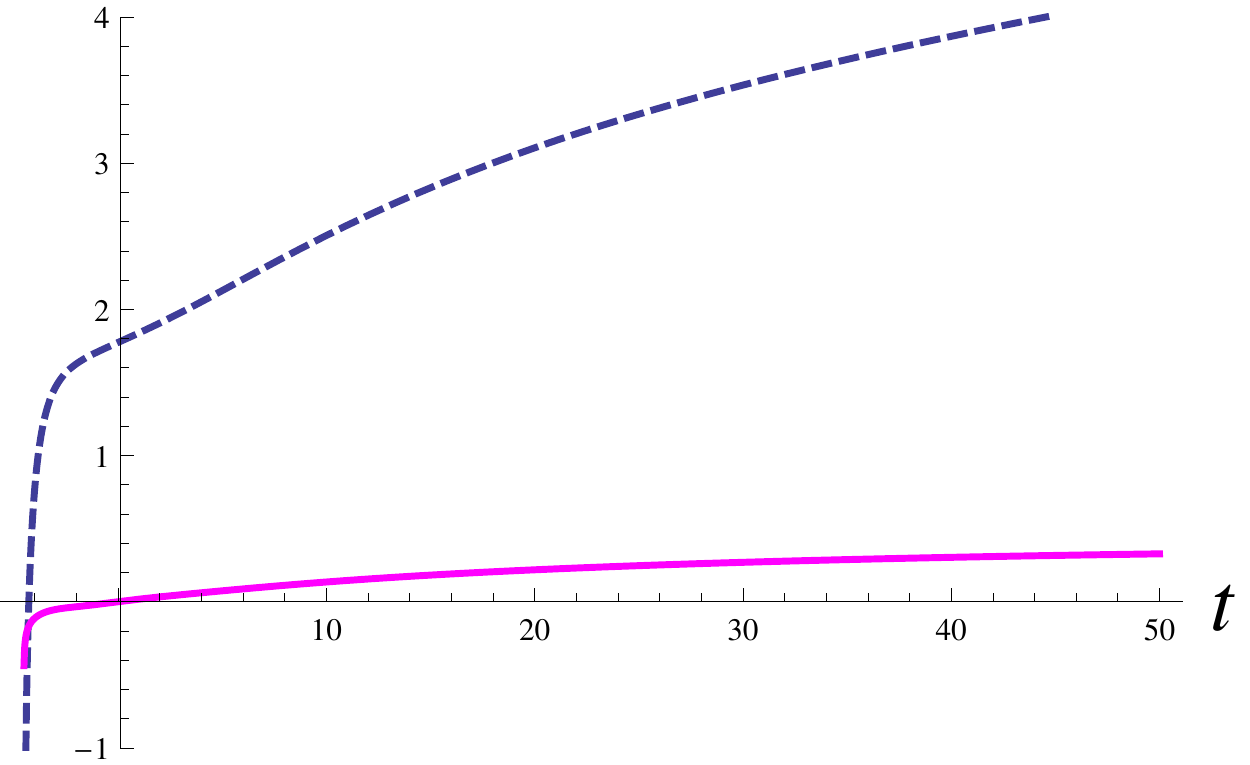} &  &  & \includegraphics[scale=0.5,keepaspectratio=true]{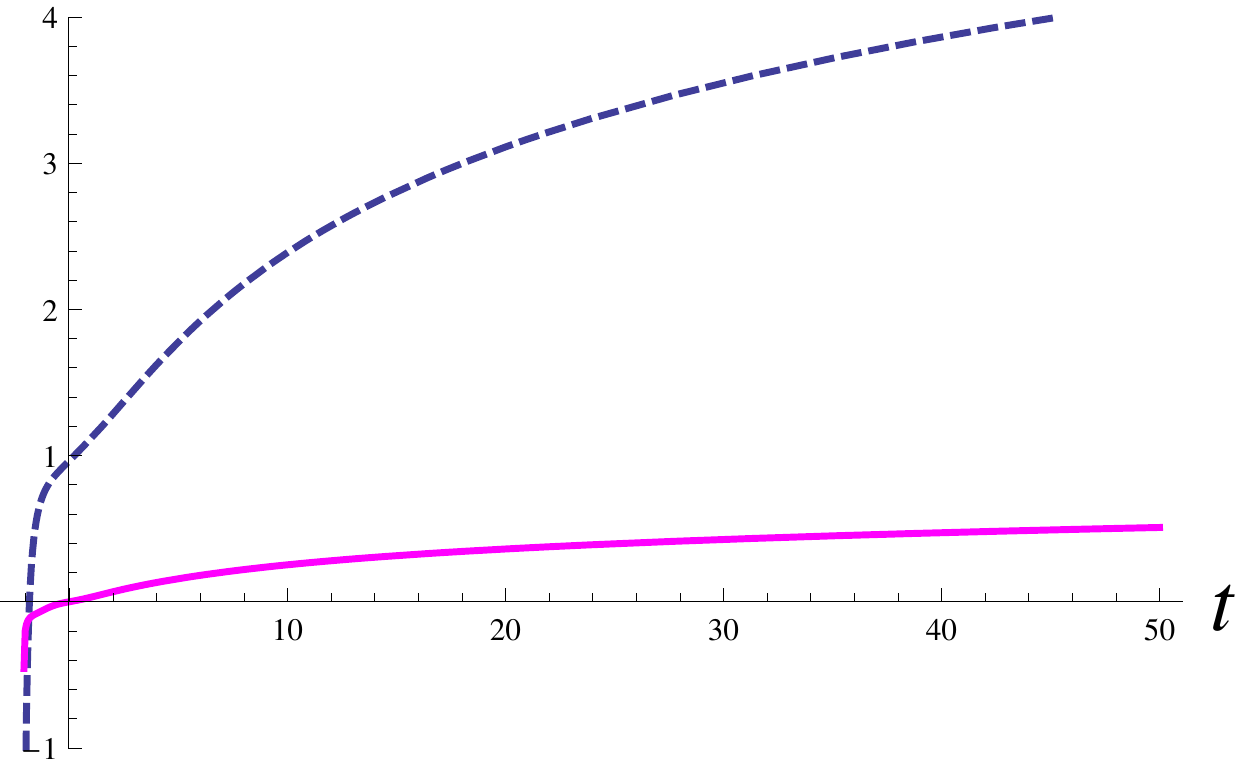}  \\
\includegraphics[scale=0.5,keepaspectratio=true]{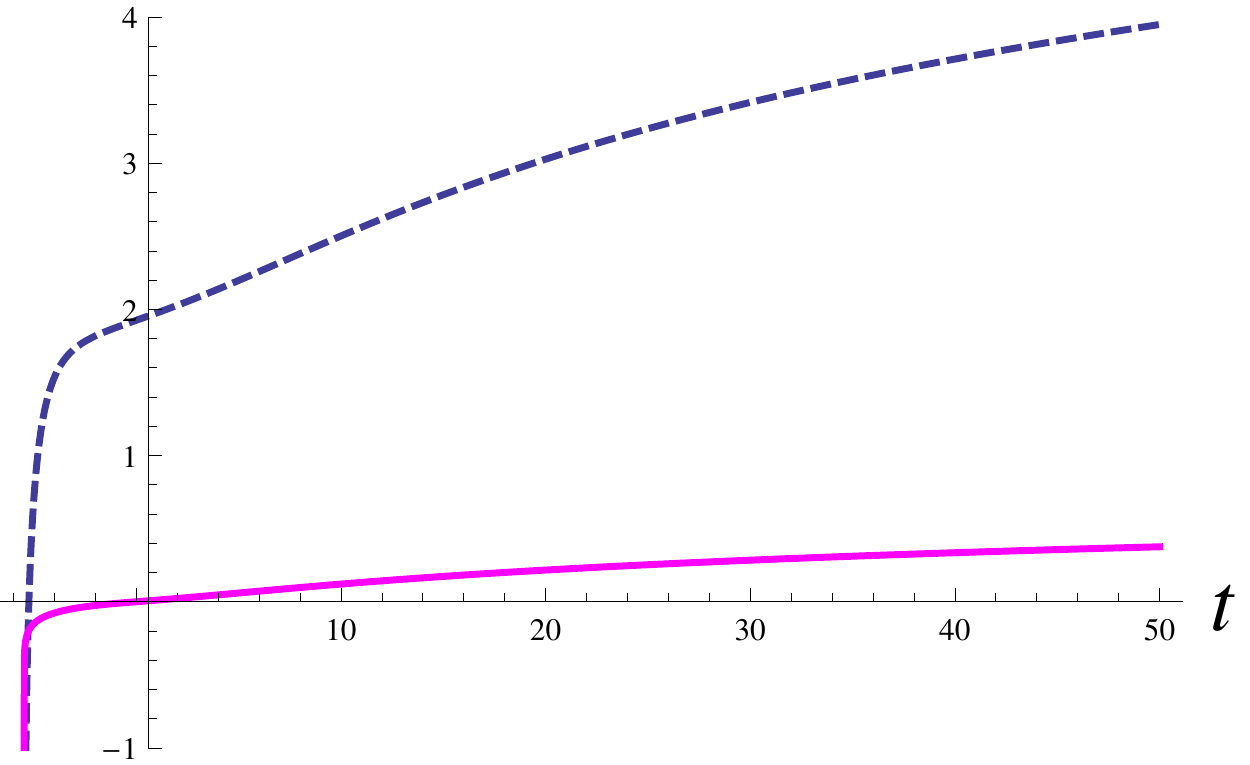} &  &  & \includegraphics[scale=0.5,keepaspectratio=true]{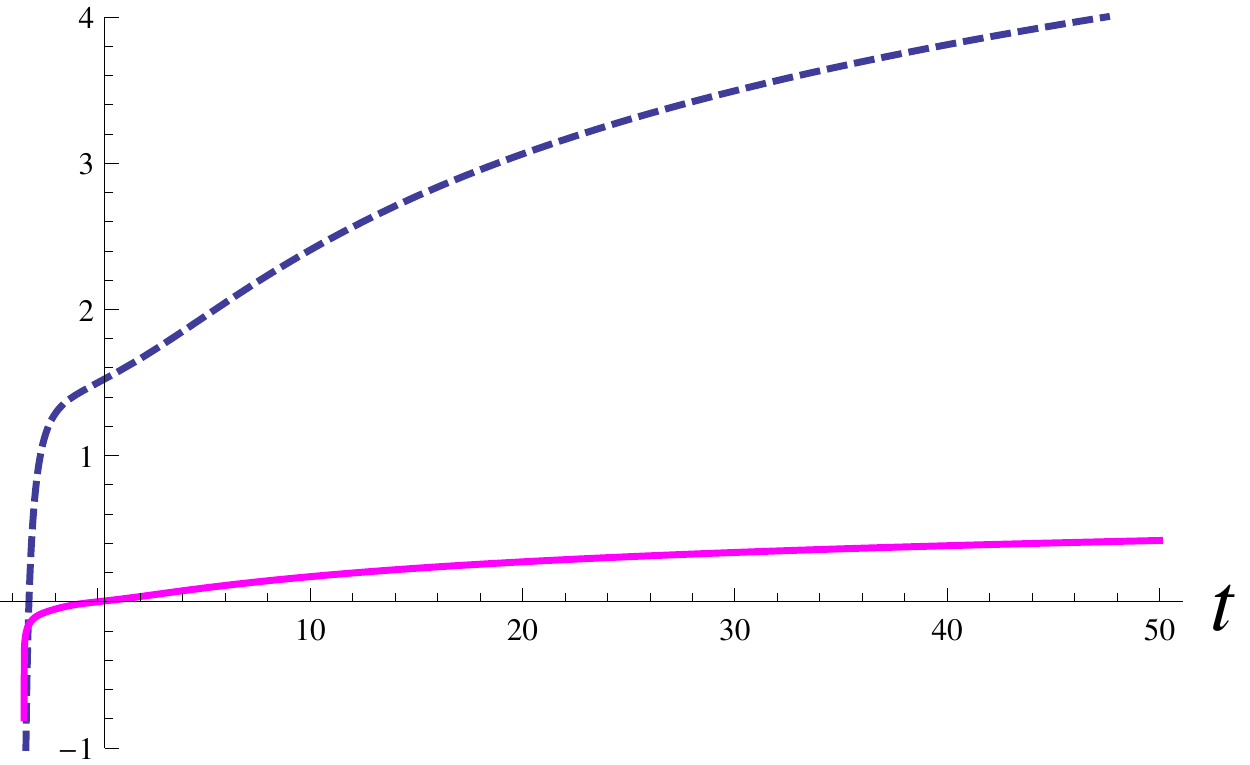}  
\end{tabular}
}
\end{center}
\caption{{\it The logarithm of the inverse Hubble scale $\log \frac{1}{H}$ (dashed blue) versus the logarithm of the KK radius $\log \rho^{1/2} \equiv \log \sqrt[6]{\textrm{Im}(U_{1})\textrm{Im}(U_{2})\textrm{Im}(U_{3})}$ (magenta). The different solutions are listed as in figure 1. Phenomenologically one would like to achieve separation between these two scales through $\frac{1}{H}\gg\rho^{1/2}$. Please note that all the quantities are given in Planck units.}}
\label{Plots_H}
\end{figure}
By performing the aforementioned choice of initial conditions, one puts the system in a slow-roll regime at $t=0$ and lets the system evolve in time to see how long it actually stays in this
phase of accelerated expansion before decaying. We have solved the coupled differential equations \eqref{EOMs} for our system in all the cases presented in tables~\ref{table:Sol_1&2}--\ref{table:Sol_3&4} 
by means of numerical methods. The corresponding resulting plots are collected in figures~\ref{Plots_addot} and ~\ref{Plots_H}. 

\begin{figure}[h!]
\begin{center}
\scalebox{0.9}[0.9]{
\begin{tabular}{ccc}
\hspace{-10mm}
\includegraphics[scale=0.45,keepaspectratio=true]{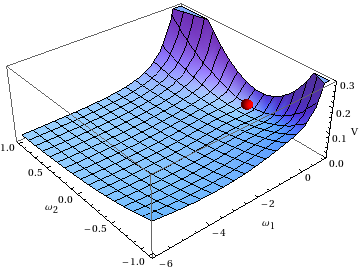} & \includegraphics[scale=0.45,keepaspectratio=true]{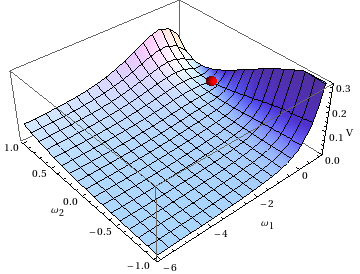} 
& \includegraphics[scale=0.45,keepaspectratio=true]{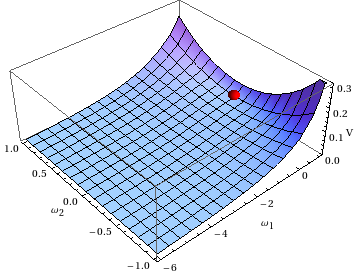}\\
\includegraphics[scale=0.45,keepaspectratio=true]{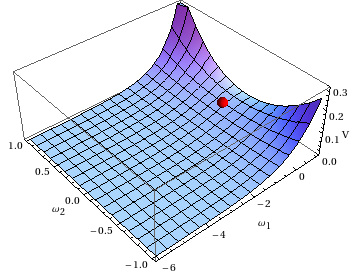} & \includegraphics[scale=0.45,keepaspectratio=true]{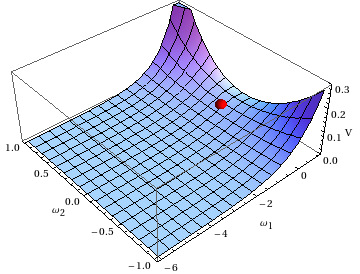} 
& \includegraphics[scale=0.45,keepaspectratio=true]{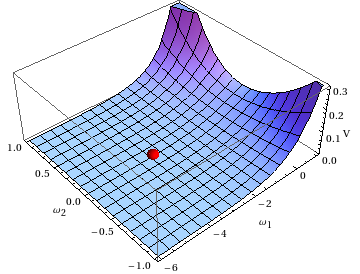}  \\
\end{tabular}
}
\end{center}
\caption{{\it 
The above images illustrate the profile of the potential of solution~$3$ in table~\ref{table:eps_eta} in the time-dependent two-dimensional field subspace given by the direction of rolling $\phi_{\epsilon}^{I}$ as defined in \eqref{rolling_dir} and the eigenstate of the mass matrix corresponding to the lowest eigenvalue that remains. 
The three top images represent the potential during the early times of the acceleration phase; $t=-3.04,\ -3,\ -2.8$, respectively.
The three lower images depict the potential at later times; the origin $t=0$, at maximum acceleration $t=0.58$, and at the end of the acceleration phase $t = 27.1$, respectively.}}
\label{Plots_3D}
\end{figure}

As can be seen from figure 1, all of the cases show an accelerated expansion phase and the corresponding amount of e-folds for each solution are given in table~\ref{table:e_folds}. Figure 2 compares the evolution of the Hubble radius $1/H$ and the KK radius $\rho^{1/2}$. Both increase as the solution moves out of the phase of accelerated expansion with the Hubble scale being the fastest one. In section 5 we will see that the solutions can be rescaled such that scale separation between the Hubble scale and the KK-scale is achieved throughout.
Moreover, in figure~\ref{Plots_3D}, we show explicitly in which direction one rolls during the time-evolved process in a two-dimensional slice of field space. Note that the potential in the two dimensional subspace shown is slightly time dependent due to a slow drift in the remaining dimensions of field space.
\begin{table}[h!]
\begin{center}
\scalebox{0.85}[0.90]{
\begin{tabular}{|c||c|c|c|c|c|c|c|c|c|c|}
\hline
ID & Sol.~$1$ & Sol.~$2$ & Sol.~$3$& Sol.~$4$ \\
\hline \hline
e-folds & $2.07368$ & $2.22914$ & $2.62382$ & $2.27844$\\
\hline
\end{tabular}
}
\end{center}
\caption{{\it The duration of the accelerated expansion phase for all four solutions given in e-folds, \emph{i.e.} $\log\frac{a(t_{\textrm{fin.}})}{a(t_{\textrm{in.}})}$.}.
}
\label{table:e_folds}
\end{table}

\section{Discussion}
\label{sec:discussion}

All the cosmologies presented in the previous section are obtained as valid four-dimensional supergravity solutions arising from geometric compactifications of type IIA string theory.
However, one would like to directly address the issue of their reliability as honest solutions of perturbative string theory together with the question of scale separation in order to meet other physical
requirements. All of this could be in principle be achieved in a regime in which the following conditions are met (where $R = \rho^{1/2}$ denotes the compactification radius):
\begin{itemize}
\item Large volume approximation, \emph{i.e.} $\frac{\sqrt{\a^{\prime}}}{R} \, \ll \, 1$,
\item Perturbative regime in the string coupling, \emph{i.e.} $g_{s} \, \ll \, 1$,
\item The cosmological constant is insensitive to Planckian physics, \emph{i.e.} $\frac{|V|}{M_{\textrm{Pl}}^{4}} \, \ll \, 1$,
\item The effective theory is four-dimensional, \emph{i.e.} $R \, H \, \ll \, 1$,
\item Very high flux quanta.
\end{itemize}

\subsection*{Perturbative control and separation of scales}

In the following we will use a ten dimensional string metric of the form
\be
\textrm{d} s^2 = \tau^{-2} \, \textrm{d} s_4^2 + \rho \, \textrm{d} s_6^2,
\ee
where $\tau$ and $\rho$ are the so called universal moduli. Compactifying down to four-dimensions gives rise to a four dimensional Planck mass given by $M_{\textrm{Pl}}^{2} \, = \, 
\frac{R^{6} \, \tau^{-2}}{g_{s}^{2} \, \left(\a^{\prime}\right)^{4}}$. If we want to end up in Einstein frame with $M_{\textrm{Pl}}=1$, we need to pick $\tau$ such that
\be
\left\{\begin{array}{lclc}
\rho & = & \left(\textrm{vol}_{6}\right)^{1/3} & , \\
\tau & = & e^{-\phi} \, \sqrt{\textrm{vol}_{6}} & ,
\end{array}\right.
\ee
where $\textrm{vol}_{6}$ is the internal volume, $\phi$ is the ten-dimensional dilaton. The various flux-induced terms in the scalar potential scale as \cite{Hertzberg:2007wc}
\be
\label{rho/tau}
\begin{array}{lclclc}
V_{H_{3}} \, \sim \, h_{3}^{2} \, \rho^{-3} \, \tau^{-2} & , & V_{\omega} \, \sim \, \omega^{2} \, \rho^{-1} \, \tau^{-2} & , & V_{F_{p}} \, \sim \, f_{p}^{2} \, \rho^{3-p} \, \tau^{-4} & ,
\end{array}
\ee
where $V_{F_{p}}$ represents the various contribution to the vacuum energy coming from the $p$-form field strength fluxes in type IIA, $p\,=\,0,\,2,\,4,\,6$ and $h_{3}, \, \omega$ and $f_{p}$ denote the corresponding 
flux quanta.

In particular, by scaling the universal moduli according to $\rho \, \sim \, N$ and $\tau \, \sim \, N^{\delta}$ and the flux quanta as
\be
\label{scalings}
\begin{array}{ccc}
\begin{array}{lc}
f_{0} \, \sim \, N^{\a-2} & , \\
f_{2} \, \sim \, N^{\a-1} & , \\
f_{4} \, \sim \, N^{\a} & , 
\end{array} & &
\begin{array}{lc}
f_{6} \, \sim \, N^{\a+1} & , \\
h_{3} \, \sim \, N^{\a-\delta+1} & , \\
\omega \, \sim \, N^{\a-\delta} & , 
\end{array}
\end{array}
\ee
where $N$ is a very large number and $\a$ and $\delta$ are suitable positive numbers, one finds
\be
\begin{array}{lclclclc}
\textrm{vol}_{6} \, \sim \, N^{3} & , & g_{s} \, \sim \, N^{\frac{3}{2}-\delta} & , & \frac{|V|}{M_{\textrm{Pl}}^{4}} \, \sim \, N^{2\a-4\delta-1} & , & \frac{H^2 \, R^{2}}{M_{\textrm{Pl}}^{2}} \, \sim \, N^{2\a-2\delta} & ,
\end{array}
\ee
where $|V| \sim H^2 M_{\textrm{Pl}}^{2}$. In the case of vanishing metric flux this regime was achieved in ref.~\cite{DeWolfe:2005uu}\footnote{Set $\a \, = \, 2$ and $\delta \, = \, 3$.}. In our more general setup, one can still go into a 
perturbative regime, but then an arbitrarily small internal curvature (\emph{i.e.} metric flux) would be needed in order to achieve separation of scales at the same time. Whether this is possible 
still remains to be seen\footnote{We thank Thomas Van Riet for pointing this out to us.}.

Moreover, we have so far neglected the issue of \emph{tadpole cancellation}. Whenever O$6$-planes and D$6$-branes are added to these compactifications, 
they induce the following cross-term in the scalar potential 
\be
V_{\textrm{O}6/\textrm{D}6} \, \sim \, N_{6} \, \tau^{-3} \ ,
\ee
where we must make sure that the net charge $N_{6}$ is cancelled by fluxes in the following way
\be
\label{tadpole}
N_{6} \, = \, f_{0} \, h_{3} \, + \, f_{2} \, \omega \, = \, - (\# \textrm{ fixed points}) \, + \, N_{\textrm{D}6} \ . 
\ee
The quantity $N_{6}$ is quadratic in the flux quanta, and will, for that reason, generically be very large in the supergravity limit unless one performs some careful finetuning. It is useful to distinguish bewteen three different cases:
\begin{itemize}
\item {\bf $N_{6} > 0$ :} This is unproblematic. High flux quanta is accompanied by large  $N_{6}$ and a large number of D$6$-branes.
\item {\bf $N_{6} = 0$ :} All flux quanta can be kept very high as long as they satisfy $f_{0} \, h_{3} \, + \, f_{2} \, \omega \, = \, 0$. No sources at all need to be added in this case and the corresponding
contribution inside the scalar potential is just absent.
\item {\bf $N_{6} < 0$ :} This corresponds to a net orientifold configuration. In this case one cannot make $|N_{6}|$ arbitrarily large since it is bounded by the number of fixed points of the corresponding
orientifold involution, which is typically a small number, say $\mathcal{O}(1)$.  
\end{itemize}
This last case turns out to be a bit troublesome, not only because it would require an enormous finetuning to combine very large flux quanta into $N_{6}\sim\mathcal{O}(1)$, but also because  such a 
contribution to the energy would tend to disappear in the supergravity limit. This effect would be due to the scaling behaviour for large $M_{\textrm{Pl}}$, under which the local source term 
would go to zero while all the others stay finite. This somehow suggests that orientifold planes are intrinsically flux-quantisation sensitive objects and cannot readily be seen by supergravity.

All our solutions fall in this third category and hence suffer from the same problem. In the next section, we will show that in fact any slow-roll accelerated quasi-dS background requires
$N_{6} \, < \, 0$, and hence the presence of O-planes.

\subsection*{Slow-roll accelerated solutions need O-planes}

The fact that dS solutions imply the presence of O-planes is well-known in the literature (see \emph{e.g.} how the technique introduced in ref.~\cite{Hertzberg:2007wc} was used 
to show this in four dimensions \cite{Danielsson:2009ff, Wrase:2010ew} and higher \cite{VanRiet:2011yc}). This can be seen very easily by writing $V$ as 
\be
\label{dS_O}
V \, = \, V_{H_{3}} \, + \, V_{\omega} \, + \, \sum\limits_{p}{V_{F_{p}}} \, + \, V_{\textrm{O}6/\textrm{D}6} \, = \, 
 \underbrace{- \, \frac{1}{2} \, \tau \, \partial_{\tau}V}_{= \,\, 0 } \, \underbrace{- \, \sum\limits_{p}{f_{p}^{2} \, \rho^{3-p} \, \tau^{-4}}}_{\leq \,\, 0} \, - \, \frac{1}{2} \, N_{6} \, \tau^{-3} \ ,
\ee
where $V \, \overset{!}{>} \, 0$ implies $N_{6} \, < \, 0$, \emph{i.e.} net orientifold charge.

In the case of quasi-dS backgrounds, the term in \eqref{dS_O} proportional to the equation of motion for $\tau$ no longer needs to be zero. Thus, in order to extend this result to this more general case,
we need to relate this term to the first slow-roll parameter $\epsilon_{V}$ introduced in \eqref{slow-roll_V}. One finds
\be
\frac{1}{4}\,\left(\frac{\tau\,\partial_{\tau}V}{V}\right)^{2} \, \leq \, \epsilon_{V} \ .
\ee
This implies
\be
\left(1 \, - \, \sqrt{\epsilon_{V}}\right) \, V \, \leq \, - \, \sum\limits_{p}{f_{p}^{2} \, \rho^{3-p} \, \tau^{-4}} \, - \, \frac{1}{2} \, N_{6} \, \tau^{-3} \ ,
\ee
and hence still $N_{6} \, < \, 0$ whenever $V \, > \, 0$ and $\epsilon_{V} \, < \, 1$.

\subsection*{Summarising}

By switching to the string frame\footnote{Please note that the string frame can be obtained by exchanging $\tau$ for $g_{s}$ via $\tau \, = \, \frac{1}{g_{s}} \, \rho^{3/2}$. In this frame, the string 
length $\ell_{s} \, \sim \, \sqrt{\alpha^{\prime}}$ rather than the Planck scale is kept constant. Accordingly, the Planck length $\ell_{\textrm{Pl}} \, \sim \, 1/M_{\textrm{Pl}}$ will now scale 
dynamically.}, one can relate the scaling behaviours presented in equation \eqref{rho/tau} to the following dependences on the KK radius $R$, $\a^{\prime}$ and $g_{s}$
\be
\label{alpha'/gs}
\begin{array}{lclc}
\frac{V_{H_{3}}}{M_{\textrm{Pl}}^{4}} \, \sim \, g_{s}^{2} \, \left(\frac{\sqrt{\a^{\prime}}}{R}\right)^{12} \, \left(\a^{\prime}\right)^{-2} \, h_{3}^{2} & , & 
\frac{V_{\omega}}{M_{\textrm{Pl}}^{4}} \, \sim \, g_{s}^{2} \, \left(\frac{\sqrt{\a^{\prime}}}{R}\right)^{8} \, \left(\a^{\prime}\right)^{-2} \, \omega^{2} & , \\[2mm]
\frac{V_{F_{p}}}{M_{\textrm{Pl}}^{4}} \, \sim \, g_{s}^{4} \, \left(\frac{\sqrt{\a^{\prime}}}{R}\right)^{6+2p} \, \left(\a^{\prime}\right)^{-2} \, f_{p}^{2}  & , & \textrm{ for } \,\,\,p\,=\,0,\,2,\,4,\,6 & .
\end{array}
\ee
The above scaling behaviours show that it is possible to achieve the large volume approximation and small string coupling at the same time and, in this regime, all terms in the scalar potential become very small 
compared to the Planck scale. However, in order to keep them finite, one needs to choose very high flux quanta. This is what one would correctly expect in the supergravity description, where one should not
be able to see flux quantisation.

It is reassuring to see that one can achieve scale separation from \eqref{scalings} according to
\be
\begin{array}{cccccccc}
\underbrace{\quad H^{-1}\quad }_{\sim \, N^{-\alpha+2\delta+\frac{1}{2}}} & \gg & \underbrace{\quad R\quad }_{\sim \, N^{\frac{1}{2}}} & \gg & \underbrace{\quad \ell_{s}\quad }_{\sim \, N^{0}} & \gg & 
\underbrace{\quad \ell_{\textrm{Pl}}\quad }_{\sim \, N^{-\delta}} & .
\end{array}
\ee
Assuming no 
finetuning, the gravitino mass is of the same order as the Hubble scale, \emph{i.e.} $m_{3/2} \, \sim \, H$.
If we fix $\delta=\alpha$, so that one does not have to worry about metric flux becoming small, one still has the freedom to control the magnitude of the Hubble scale by $\alpha$.  
The only trouble is $|N_{6}| \sim N^{2\alpha-\delta-1}=N^{\alpha-1}$, where we must have $\alpha>3/2$ in order for $g_s \rightarrow 0$ at large $N$. Since $N_{6}$ cannot be made arbitrarily negative, there is a limit how far one can push the scale $N$ to large values. 

A loop hole would be if there is some finetuning in order to satisfy the tadpole cancellation condition in \eqref{tadpole} with high flux numbers scaling as described in \eqref{scalings}. Then one could cancel the leading 
$N$-scaling inside $|N_{6}|$ and the Hubble scale itself. This can effectively give a much smaller cosmological constant than what the leading scaling behaviour would suggest and, in particular, it would also be much smaller than $m_{3/2}$.

The issue of quantisation, and the lower limit on $N_{6}$, is something that possibly could put the supergravity approach into doubt. Note, though, that the usual loop corrections are of order $g_{s}^{2} \, V$ \cite{Berg:2005ja}, and thus under control.

\section{Conclusions}
\label{sec:conclusions}

We have studied the problem of cosmic accelerated expansion in type IIA backgrounds with metric and gauge fluxes in the context of $\mathbb{Z}_{2} \, \times \, \mathbb{Z}_{2}$ orbifold compactifications.
With the aid of a genetic algorithm, we were able to find backgrounds with both slow-roll parameters within order of $10\%$. Subsequently, by explicitly studying the corresponding time-dependent
dynamics in these cases, we have shown that there are some backgrounds developing an accelerated expansion phase lasting for a few e-folds.

When discussing the possibility of achieving perturbative control and separation of scales, we found that there are different cases to be analysed according to the possible different choices for the 
exponents $\alpha$ and $\delta$ scaling the fluxes in \eqref{scalings}. The most promising case for achieving scale separation without having metric flux going to zero, seems to be when $\delta=\alpha$.
In order to avoid the tadpole $N_{6}$ to grow arbitrarily high (as $N^{\alpha-1}$), one would need some finetuning for cancelling this leading divergent contribution between the two different terms
appearing in \eqref{tadpole}. This same finetuning will also make the cosmological constant small compared to the Planck scale and simultaneously separate all the other scales. In addition, loop-corrections are estimated to be
under control in such a situation, since their size is roughly given by $g_{s}^{2} \, V$.

To summarise, we find it extremely encouraging that the relaxation of the requirement of time independence, opens up new possibilities to look for cosmologically interesting solutions. It is intriguing to note that the quasi-dS solutions we have found exhibit a sufficient number of e-foldings to be relevant for late time dark energy. Furthermore, through an appropriate choice of scaling we can tune down quantities such as the gravitino mass and the cosmological constant to quite small values, and achieve scale separation with respect to other important scales. With finetuning one should be able to separate these two scales and make the cosmological constant even smaller. It remains to be seen whether it is possible to find examples with phenomenologically interesting values.

%
%

\section*{Acknowledgments}

We would like to thank David Berman, Mariana Gra$\tilde{\textrm{n}}$a, Adolfo Guarino, Diego Marqu\'es and Thomas Van Riet for interesting and stimulating discussions. 
The work of the authors is supported by the Swedish Research Council (VR), and the G\"oran Gustafsson Foundation.

\newpage

%
%

\appendix

\section{Tables of flux values}
\label{sec:app}
\begin{table}[h!]
\begin{center}
\begin{tabular}{|c||c|c|}
\hline
& Sol.~$1$& Sol.~$2$\\
\hline
$a_0$& $0.0286315$& $0.25972$\\
\hline
$a_1^{(i)}$& $\begin{array}{ccc}-0.435346&0.0148263&-0.198035\end{array}$& $\begin{array}{ccc}0.741984&-0.0314638&0.505144\end{array}$\\
\hline
$a_2^{(i)}$& $\begin{array}{ccc}0.294943&-0.337634&-0.81373\end{array}$& $\begin{array}{ccc}-0.621304&0.651788&1.58042\end{array}$\\
\hline
$a_3$& $1.77098$& $3.96483$\\
\hline
$b_0$& $-0.568105$& $-1.01951$\\
\hline
$b_1^{(i)}$& $\begin{array}{ccc}0.705827&-0.240703&-1.38994\end{array}$& $\begin{array}{ccc}-1.37824&0.110598&3.16408\end{array}$\\
\hline
$c_0^{(i)}$& $\begin{array}{ccc}0.717192&-0.020487&0.428282\end{array}$& $\begin{array}{ccc}1.38446&-0.661604&1.54381\end{array}$\\
\hline
$c_1^{(ij)}$& $\begin{array}{ccc}0.119802&0.0408551&0.235917\\0.736266&0.251083&-1.44988\\-0.292539&0.0997625&-0.576077\end{array}$& $\begin{array}{ccc}-0.164483&-0.0131991&-0.37761\\-1.5181&-0.121821&3.48516\\0.740001&-0.059382&1.69885\end{array}$\\
\hline
\end{tabular}
\end{center}
\caption{{\it The flux values identifying Sol.~$1$ and $2$. All the scalars are sitting at the origin of moduli space.}}
\label{table:Sol_1&2}
\end{table}
\begin{table}[h!]
\begin{center}
\begin{tabular}{|c||c|c|}
\hline
& Sol.~$3$& Sol.~$4$\\
\hline
$a_0$& $0.00300432$& $0.00880628$\\
\hline
$a_1^{(i)}$& $\begin{array}{ccc}-0.0352252&0.011383&0.0242228\end{array}$& $\begin{array}{ccc}-0.0879238&0.0110747&-0.0788949\end{array}$\\
\hline
$a_2^{(i)}$& $\begin{array}{ccc}0.276756&-0.58295&0.0721174\end{array}$& $\begin{array}{ccc}0.656355&-0.26349&-0.835823\end{array}$\\
\hline
$a_3$& $-1.68959$& $-2.43669$\\
\hline
$b_0$& $0.37311$& $0.965429$\\
\hline
$b_1^{(i)}$& $\begin{array}{ccc}0.305757&-0.222611&0.216332\end{array}$& $\begin{array}{ccc}0.952404&-0.136589&-1.39332\end{array}$\\
\hline
$c_0^{(i)}$& $\begin{array}{ccc}-0.46172&0.234389&-1.00233\end{array}$& $\begin{array}{ccc}-0.886049&0.379069&-0.834596\end{array}$\\
\hline
$c_1^{(ij)}$& $\begin{array}{ccc}-0.941403&-0.685405&0.66607\\-0.949435&-0.691253&-0.671753\\-0.302185&0.220011&0.213804\end{array}$& $\begin{array}{ccc}0.340021&0.048764&0.497434\\1.0941&0.15691&-1.60062\\-0.388406&0.0557031&-0.568219\end{array}$\\
\hline
\end{tabular}
\end{center}
\caption{{\it The flux values identifying Sol.~$3$ and $4$. All the scalars are sitting at the origin of moduli space.}}
\label{table:Sol_3&4}
\end{table}
%

%
%

\small

\clearpage

\bibliography{references}
\bibliographystyle{utphys}

\end{document}